\documentclass[preprint,aps,nofootinbib]{revtex4}
\usepackage{amsmath}
\usepackage{latexsym}
\usepackage[pdfauthor={Alexander Dolgov, Diego N. Pelliccia},pdftitle=
{Electric asymmetry of the universe},pdfsubject=
{electric asymmetric universe}, pdfkeywords={galactic magnetic fields, 
black holes, electric asymmetry, photon mass, intergalactic plasma, 
cosmic acceleration, dark energy}]{hyperref}
\newcommand{\be}{\begin{equation}}
\newcommand{\ee}{\end{equation}}
\newcommand{\ba}{\begin{eqnarray}}
\newcommand{\ea}{\end{eqnarray}}
\newcommand{\bi}{\bibitem}
\newcommand{\mg}{m_\gamma}
\newcommand{\lamg}{\lambda_\gamma}
\newcommand{\rar}{\rightarrow}

\newcommand{\bc}{\begin{center}}
\newcommand{\ec}{\end{center}}
\def\cstok#1{\leavevmode\thinspace\hbox{\vrule\vtop{\vbox{\hrule\kern1pt
\hbox{\vphantom{\tt/}\thinspace{\tt#1}\thinspace}}
\kern1pt\hrule}\vrule}\thinspace}
\begin{document}
\title{Photon mass and electrogenesis}
\author{Alexander Dolgov$^{1,2,3}$
\thanks{Electronic address: dolgov@fe.infn.it},
Diego N. Pelliccia$^{1,2}$
\thanks{Electronic address: diego.pelliccia@fe.infn.it}}
\affiliation{ ${ }^{1}$Istituto Nazionale di
Fisica Nucleare, Sezione di Ferrara,\\
Via Saragat 1, 44100 Ferrara, Italy\\
${ }^{2}$Universit\`a di Ferrara,
Dipartimento di Fisica,\\
Via Saragat 1, 44100 Ferrara, Italy\\
${ }^{3}$ITEP,
Bol. Cheremushkinskaya 25, Moscow 113259 Russia\\
}

\vspace{2cm}
\begin{abstract}

We show that if photon possesses a tiny but non-vanishing mass the
universe cannot be electrically neutral. Cosmological electric 
asymmetry could be generated either at an early stage by different 
evaporation rates of primordial black holes with respect to 
positively and negatively charged particles or by predominant 
capture of protons in comparison to electrons by heavy galactic 
black holes in contemporary universe. An impact of this phenomenon on 
the generation of large scale magnetic fields and on the universe 
acceleration is considered.

\end{abstract}

\maketitle

The universe is known to be asymmetric with respect to
particles and antiparticles possessing, at least locally,
non-zero baryonic and, possibly, leptonic charge density.
On the other hand, it is commonly believed that cosmological
electric asymmetry must be identically zero. The observational 
data allow for some small cosmological electric  
asymmetry~\cite{constraints}, but theoretically it is very
difficult or maybe even impossible to generate it without 
modification of the usual Maxwell electrodynamics.
There are two mutually related reasons for the absence of any 
cosmological electric asymmetry: electric current conservation 
and vanishing of the photon mass. 

Attempts to break electric current conservation 
theoretically~\cite{ch-okun,ch-theor} or to observe it 
experimentally~\cite{ch-exp,pdg} have a long history.
If photon mass is zero, then the Maxwell equations automatically 
imply current conservation. Indeed from
\be
\label{max}
\nabla_{\mu}F^{\mu\nu} = 4\pi J^\nu
\ee
follows 
\be
\nabla_{\nu} J^\nu = \frac{1}{\sqrt{-g}}
\partial_\nu \left(\sqrt{-g} J^\nu\right) =0,
\label{current-cons}
\ee
because the divergence of the l.h.s. of equation (\ref{max})
is identically zero due to the antisymmetry of the Maxwell field 
tensor, $F^{\mu\nu}$. Here $\nabla$ is the covariant derivative
in background gravitational field.

The only way to break current conservation is to modify the
Maxwell equations. It can be done, in particular, 
by an introduction of non-zero photon mass. After the
two-dimensional example discovered by Schwinger \cite{Scwhinger}, 
several mechanisms have been proposed to generate photon mass in
four dimensions. The photon mass could be created as a result
of spontaneous breaking of the gauge $U(1)$-symmetry or,
possibly, topologically \cite{Jackiw} in analogy with the original
suggestion~\cite{Scwhinger}. Here we will not go into theoretical
justification for the photon mass but simply assume that it 
exists and consider the cosmological implications of this assumption.

In the case of non-zero photon mass, instead of eq. (\ref{max}), 
we would have the Proca equation:
\be
\label{proca}
\nabla_{\mu}F^{\mu\nu} +\mg^{2} A^\nu = 4\pi J^\nu.
\ee
The current may be non-conserved and new longitudinal degrees of
freedom of the vector field $A_\mu$ would appear:
\be
\nabla_{\nu} A^\nu = \frac{4\pi}{\mg^{2}} \nabla_{\nu} J^\nu \, .
\label{current-non-cons}
\ee
However, because of the extreme smallness of the experimentally allowed
value of the photon mass the effects of current non-conservation
lead to disastrous infrared problems in the theory~\cite{ch-okun}.
The photon mass is best bounded from above by astronomical observations 
of magnetic fields of celestial bodies or galactic magnetic fields,
for a review see ref.~\cite{m-gamma-rev}.  A robust and quite strong
limit is found from the measurement of the Jupiter magnetic field by the Pioneer-10
mission~\cite{jupiter}, which gives
\be
\mg < 6\cdot 10^{-16}\,\,{\rm eV}\,\,\, {\rm or}\,\,\,
\lamg > 3\cdot10^{10}\,\,{\rm cm},  
\label{mg-jupiter} 
\ee
where $\lamg\equiv 1/\mg$ is the Compton wavelength of the photon.

A much stronger result is obtained from the fact that there exist
galactic magnetic fields coherent on the galactic size of a few
kpc~\cite{chibisov}:
\be
m_{\gamma} < 10^{-27}\,\,eV\,\,\, {\rm or}\,\,\,
\lamg > 10^{22}\,\,{\rm cm}.  
\label{mg-gal}
\ee
However, it was argued in ref.~\cite{dvali-gruz} that if photon acquires its 
mass through spontaneous breaking of the gauge $U(1)$-symmetry, the 
bound (\ref{mg-gal}) may be invalid, while the bound (\ref{mg-jupiter})
still remains. On the other hand, as argued in ref [2], spontaneous breaking of the electro-magnetic $U(1)$-symmetry leads to the existence of a new light scalar, which is excluded by experiments.

Anyhow the photon mass is very strongly bounded from above and 
such a tiny value of $\mg$ would lead to a catastrophic 
emission of the longitudinal photons in the case of non-conserved
current~\cite{ch-okun} .  The probability of 
this process is proportional to a power of $(\nabla J\, E/\mg)^2$,
where $E$ is the characteristic energy in the reaction. Moreover,
the diagrams with virtual longitudinal photons demonstrate the same
kind of mass singularity and, as argued in ref.~\cite{ch-okun}, it 
can create a very strong, exponential, suppression of charge 
non-conserving processes.

Due to this complications and pathological behavior of theory
of very light vector field interacting with non-conserved
current, we will consider, in what follows, only the case when  
current conservation condition (\ref{current-cons})
is strictly fulfilled.

Still the current conservation does not prevent from generation of
cosmological electric asymmetry if photons are massive. Electric charge 
could be captured by a black hole (BH) and, if $m_\gamma \neq 0$,  
the Coulomb field or, more precisely, the Yukawa field would vanish
outside BH. In this sense electric charge disappears inside BH
without any trace. This effect is discussed in several
papers~\cite{vilenkin,ll,dmt} and is in a sharp contrast with
massless electrodynamics. For massless photons a charged BH creates
the usual Coulomb field, $A_t =Q/r$, in outer space. Thus electric
charge never disappears from our world. If $m_\gamma \neq 0$ one
may naively expect that $A_t =Qe^{-m_\gamma r}/r$, but this is very 
wrong and $A_t =0$ for any arbitrarily small $m_\gamma$. As we see
in what follows, if a charged particle is outside BH horizon, the
Yukawa potential behaves as $A_t = Q e^{-m_\gamma r} (R_c-R_g)/rR_c$,
where $R_g$ is the Schwarzschild radius of BH and $R_c>r_g$ is the
position of the charged particle. Thus we see that electric charge 
creates weaker and weaker field when it approaches BH. In this sense
we can speak about an effective electric charge nonconservation.  
This phenomenon makes possible the generation of cosmological electric asymmetry.

Cosmology of electrically charged universe was considered in a 
number of papers~\cite{ch-cosm}, both with massless and massive photons.
If $\mg =0$, the uniformly charged universe, with electric charge
density $J_t\equiv \sigma$, cannot be homogeneous and
isotropic. Indeed from the Gauss law in the
Friedman-Robertson-Walker (FRW) metric follows:
\be{ 
\frac{1}{\sqrt{-g}} \partial_\mu \left( \sqrt{-g} F^{\mu 0}\right) =
{\rm div}\, E = 4\pi \sigma\, .
\label{gauss}
}\ee
This implies that the vector of electric field, $E$, should be a rising function of
coordinates:
\be
E \sim x
\label{E-of-x}
\ee
The situation would be drastically different if photons are massive.
Instead of eq. (\ref{gauss}) we would have:
\be{ 
{\rm div}\, E +\mg^2 A_t = 4\pi \sigma
\label{m-gauss}
}\ee
where $A_t$ is the time component of the vector potential. 
In this case an isotropic homogeneous solution evidently exists and takes
the form:
\be 
A_t (t) = 4 \pi \sigma /m_\gamma^2
\label{A-of-t}
\ee
with zero electric field, $ E = 0$, and 
vanishing space components of the 
vector potential, $A_i =0$. Because of current conservation
(\ref{current-cons}), the charge density in the homogeneous case 
 drops as $\sigma \sim 1/a^3(t)$,
where $a(t)$ is the cosmological scale factor. Correspondingly
$A(t)$ (\ref{A-of-t}) drops down by the same law.

The energy-momentum tensor corresponding to solution (\ref{A-of-t})
is singular at $\mg \rar 0$ because of the terms:
\be
T_{\mu\nu} =  -2 {m_\gamma^2 A_\mu A_\nu -
 g_{\mu\nu}m_\gamma^2 A^\alpha A_\alpha } + ...
\label{T-of-m}
\ee
The complete expression for $T{\mu\nu}$ is presented below, eq. (\ref{emt}),
where the terms, which ensure positive definiteness of the energy
density are included.
 
Because of the singularity $1/\mg^2$ in the energy density a
spontaneous generation of the electric field may be possible when
formation of domains with chaotically distributed electric fields,
behaving according to eq. (\ref{E-of-x}),  
which has a smooth limit to $\mg=0$, would be energetically 
more favorable than the homogeneous but singular
in $\mg$ solution (\ref{A-of-t}).  

In analogy with baryogenesis, for a generation of cosmological
electric asymmetry three Sakharov's conditions~\cite{sakharov} are 
expected to be fulfilled: 
1) non-conservation
of electric current (remember, however, that we have
assumed that electric current is conserved), 
2)~breaking of C and CP, and 3) deviations from
thermal equilibrium .
There have been considered scenarios~\cite{lang-pi,ad-js} 
where the electromagnetic $U(1)$
symmetry was broken at the early cosmological stage and restored
at lower temperatures (in contrast to the electroweak theory, where
symmetry is broken at low $T$). However,
if the symmetry was broken spontaneously, then after its restoration
the net electric charge would disappear due to the release of the
compensating charge from the vacuum~\cite{ad-js}. Still, the electric
currents generated in the process of neutralization could be traceable
through magnetic fields generated at this epoch. Such fields could be
the seeds of the observed magnetic fields in galaxies~\cite{ad-js}.

In this work we will consider only the case of formally conserved 
current, when the operator equation (\ref{current-cons}) is fulfilled.
We assume also that the mass of the photon is small but non-zero. In such
a case an electric charge asymmetry of the universe
must be created because of an electrically asymmetric evaporation of 
light black holes or by asymmetric
charge capture by heavy black holes, which mimic charge non-conservation.

The process of electrogenesis by black holes is exactly analogous to
baryogenesis by black holes with conserved baryonic charge. It was 
discussed in the original papers~\cite{bs-bh} and in the
review~\cite{bs-rev}, where it is explained in detail that the
condition of charge nonconservation is unnecessary for this 
mechanism.

As is well known, in the standard theory black holes (BH) may have 
non-vanishing gravitational field related to their mass and rotation 
and Coulomb field created by their electric charge.
They are indistinguishable otherwise. However, if photon is, even
a little, massive, Coulomb field outside BH would disappear without any
trace. In other words,
black holes do not have electric hairs if the gauge boson 
(photon) has non-vanishing mass.

For massive photons in flat space-time the
Coulomb potential 
changes into the Yukawa one:
{\be{
A_t = \frac{Q}{r} \rar \frac{Q\,e^{-m_\gamma r}}{r},
}
\label{coulomb-flat}
\ee}
but in Schwarzschild space-time outside a charged BH
the normal Yukawa (or, if $m_\gamma r \ll 1 $,  Coulomb) 
potential becomes identically zero
for any small $\mg\neq 0$~\cite{ll, vilenkin, dmt}:
\be
A_t = \frac{Q}{r} \rar A_t = 0
\label{coulomb-BH}
\ee
This can be seen from the solution of the Proca equation in
the Schwarzschild metric. In a simple spherically symmetric 
case when a charged shell with radius $R_c$ surrounds a BH
with radius $R_g$ the Proca equation for the Coulomb potential
takes the form
\begin{eqnarray}
\frac{1}{r^2}\left( r^2 A'_t \right)' - \frac{\mg^2 r}{ r - R_g}\, A_t
=0,  
\label{eqout}
\end{eqnarray}
where prime means derivative with respect to  ${ r}$. 

Introducing the effective charge
${ q = r\,A_t }$ and the new variable ${ y = 2\mg (r-R_g)}$
we find
\begin{eqnarray}
\frac{d^{2}q}{dy^{2}} - \left( \frac{1}{4} + \frac{\mu}{2y}\right)\, q =0,
\label{q''2}
\end{eqnarray}
where $\mu = m_{\gamma} R_g$. 

This is the Whittaker equation with the solution 
expressed in terms of confluent hypergeometric functions~\cite{gr}:
\begin{eqnarray}
{{
q(y)}} &=&{{ C\,y\, e^{-y/2}\,\Phi (1+\mu/2,2,y)
+B\,y\, e^{-y/2}\,\Psi (1+\mu/2,2,y),
}}
\label{y-sol2}
\end{eqnarray}
The coefficients $C$ and $B$ are found from the condition of vanishing
of the field at infinity and of absence of singularity on the horizon,
$y=0$. Otherwise the energy would be infinitely big. One can check that
when the radius of the charged shell tends to the gravitational radius,
${ R_c \rar R_g}$, 
\be
\label{q-eff}
q \sim q_0\,\frac{R_c-R_g}{R_c}\,\Gamma(1+\mu/2),
\ee
where $\Gamma(z)$ is the usual gamma-function and
$q_0 $ is the electric charge of the shell in flat space-time.
We see  that the electric field vanishes as the charged shell approaches
the gravitational radius and  disappears completely when
the charged shell is swallowed by the black hole.
We obtain thus an effective non-conservation of electric charge, 
despite formal current 
conservation. This is a well known phenomenon of breaking of global symmetries
by black holes.
Such vanishing of electric field outside BH was demonstrated in
refs.~\cite{vilenkin,ll,dmt} for a pointlike charge particle 
approaching BH.

Seemingly the transition from $\mg \neq 0$ to $\mg =0$ is discontinuous
in presence of a black hole. If ${ m_\gamma = 0}$,
${ q = const,}$ while if  ${ m_\gamma \neq 0}$, 
$ q \sim (R_c-R_g)/R_g $. On very large distances the field disappears 
anyhow because of the exponential Yukawa cutoff, $\sim \exp (-\mg r )$, but
for ${ m_\gamma r <1}$ the field would be effectively the Coulomb one with
the diminishing charge, while the shell approaches the gravitational radius.
The mass discontinuity is indeed present but the characteristic time
of disappearance of the Coulomb/Yukawa field when $R_c\rar R_g$ is 
inversely proportional to $\mg$~\cite{Pawl}. 
Characteristic time of vanishing of the electric (or magnetic) field,
$ \tau_C \sim 1/m_\gamma$, would be $\sim 1$ sec if the 
Jupiter bound (\ref{mg-jupiter}) is saturated and much longer,   
$\geq 10^4$ years, if the
galactic bound (\ref{mg-gal}) is true.

A very interesting possibility would be open if
there exists a non-minimal coupling of electromagnetic 
vector-potential to gravity:
\be
{\cal L}_{R} = (\xi/2) R A^{\mu} A_{\mu},
\label{L-R}
\ee
where $R$ is the curvature scalar and $\xi$ is a constant. In what follows we assume that $\xi\, R>0$. The case of negative $\xi\, R$ corresponds to Higgs type instability and, though it may be interesting, is not considered here. According to the
Einstein equations $R$ is expressed through the trace of the energy-momentum
tensor:
\be
R = 8\pi T /m_{Pl}^2 
\label{R}
\ee
In cosmological situation during the matter dominated stage
$R = 4/3t^2$, where $t$ is the universe age. During the radiation dominated
stage $R \approx 0$ up to corrections related to the particle masses and the 
conformal anomaly. 

In what follows, the generation of electric charge asymmetry
by BH in galaxies is considered and to this end it is 
instructive to know the values of the effective photon mass there.
For the average galactic mass density, with dark matter included,
$\rho \approx 10^{-24} $ g/cm$^3$ and we find
\be
1/R^{-1/2} \approx 10^7\,\,\,{\rm years}\, .
\label{R-gal}
\ee
The mass density in the galactic center near the central black hole
may be much higher~\cite{gondolo-silk} and the effective time 
of ``consumption'' of electric field by 
a black hole would be much shorter:
\be
\tau_C \sim 1/(\xi\,R^{-1/2}) \approx 1/\xi^{1/2}\,\,\,{\rm year}\, .
\label{R-gal-cntr}
\ee
Such a large value of the effective mass of photon does not contradict
 bound (\ref{mg-gal}) because the average effective photon
mass in a galaxy, (\ref{R-gal}), is much smaller. 

To describe the cosmological consequences of a non-vanishing 
photon mass and/or of the non-minimal coupling (\ref{L-R}) let
us consider the model with a charged scalar field $\varphi$
described by the Lagrangian:
\ba
\label{lagtot}
\mathcal{L}_{tot}& = &\sqrt{g}\Big\{ -\frac{R}{16\pi\,m_{Pl}^{2}}+
\frac{\xi}{2}\,R\,
A^{\mu}\,A_{\mu}+\frac{1}{2}\,m^{2}_{\gamma}\,A^{\mu}\,A_{\mu}-\frac{1}{4}F_{\mu\nu}\,F^{\mu\nu}\nonumber\\
& &-m_{\varphi}^{2}\,|\varphi|^{2}+\left[\left(\nabla_{\mu}\,+i\,e\,A_{\mu}\right)\,
\varphi^{*}\right]\,\left[\left(\nabla_{\nu}-i\,e\,A_{\nu}\right)\,
\varphi\,g^{\mu\nu}\right]\,\Big\}.
\ea
The energy-momentum tensor for such a theory has the form:
\ba
\label{emt}
T^{\mu\nu}&&=\frac{1}{4}\,g^{\mu\nu}\,F_{\sigma\tau}\,F^{\sigma\tau}-F^{\mu\sigma}\,
F^{\nu}_{\phantom{\nu}\sigma}\nonumber\\
[1pt]&&+\xi\left[ \left( R^{\mu\nu}-\frac{1}{2}\,g^{\mu\nu}\,R\right)\,A^{2}+
\left( g^{\mu\nu}\,\nabla^{2}-\nabla^{\mu}\,\nabla^{\nu}\right)\,
A^{2}-R\,A^{\mu}\,A^{\nu}\right]\nonumber\\
[1pt]&&+g^{\mu\nu}\left[ m_{\varphi}^{2}\,|\varphi|^{2}-(\nabla_{\tau}\,\varphi^{*})\,
\nabla^{\tau}\,\varphi+i\,e\,(\varphi\,\nabla_{\tau}\,\varphi^{*}-\varphi^{*}\,
\nabla_{\tau}\,\varphi)\,A^{\tau}-e^{2}\,A^{2}\,|\varphi|^{2} -\frac{1}{2}m^{2}_{\gamma}A^{\varrho}\,A_{\varrho} \right]\nonumber\\
[1pt]
&&-2\left[(\nabla^{\mu}\,\varphi^{*})\,\nabla^{\nu}\,
\varphi+e^{2}|\varphi|^{2}\,A^{\mu}\,A^{\nu}-\frac{1}{2}\,m^{2}_{\gamma}\,A^{\mu}A^{\nu}+i\,e\,\left(A^{\mu}\,
\varphi^{*}\,\nabla^{\nu}\,\varphi-(\nabla^{\mu}\,\varphi^{*})\,A^{\nu}\,\varphi\right)\right]
\ea
Electric current of the scalar $\varphi$ is equal to:
\be
J^{\mu}=-2e^{2}\,|\varphi|^{2}\,A^{\mu}+i\,e\,\left[\varphi\,\nabla^{\mu}\,
\varphi^{*}-\varphi^{*}\,\nabla^{\mu}\,\varphi\right].
\label{J}
\ee 
Both $T^{\mu\nu}$ and $J^\mu$ are covariantly conserved
by virtue of the equations of motion:  
\ba
\label{einstein}
&&R_{\mu\nu}-\frac{1}{2}\,g_{\mu\nu}\,R=-\frac{8\pi}{m_{Pl}^{2}}\,T_{\mu\nu},
\\%\ee
%\be
\label{curvedA}
&&\nabla_{\mu}\,F^{\mu\nu}+\xi\,R\,A^{\nu}+m^{2}_{\gamma}\,A^{\nu}+2\,e^{2}\,|\varphi|^{2}\,A^{\nu}+i\,e\,
[\varphi^{*}\,\nabla^{\nu}\,\varphi-\varphi\,\nabla^{\nu}\,\varphi^{*}]=0,
\\%\ee
%\be
\label{scalar}
&&\nabla^{2}\,\varphi+m_{\varphi}^{2}\,\varphi-e^{2}\,A_{\nu}\,A^{\nu}\,\varphi-i\,e\,
(\nabla_{\nu}\,A^{\nu})\,\varphi - 2\,i\,e\,A^{\nu}\,\nabla_{\nu}\,\varphi=0.
\ea

The homogeneous and isotropic cosmological solution with a non-vanishing 
electric charge density is easy to find. The time component of the
vector potential is given  by eq. (\ref{A-of-t}) and
the electric part of the energy and pressure densities are
\be
\rho_{el} \approx p_{el} \approx \frac{\sigma^2}{2\mg^2}\, .
\label{rho-el}
\ee
Strictly speaking this result is true for fermions or for very small $e^{2}\,\langle \varphi^{2}\rangle$.
It is the so called maximally rigid equation of state when the
speed of sound is equal to the speed of light and, if the universe
was dominated by the electric contribution to $T_{\mu\nu}$ it
would expand as $a(t) \sim t^{1/6}$.

As we have already mentioned above,
there are two simple mechanisms of 
electrogenesis by black holes. The first one is realized by charge 
asymmetric evaporation of primordial BH in the early
universe and is analogous to baryogenesis by BH evaporation~\cite{bs-bh}. 
Let us assume that BH emits e.g. a neutral particle, $X^0$, which 
decays near horizon into 
two charged particles, $X^0 \rar A^+B^-$ and $\bar X^0 \rar A^-B^+$ with
different decay rates. Such decays could be, e.g. $X^0 \rar t+ \bar u$,
where $t$ is a heavy quark and $\bar u$ is a light antiquark. 
The back-capture of $A$ and $B$ by BH would be different
if $m_A\neq m_B$ and thus 
electric charge would be accumulated in outer space.
Such processes may go even if the photon mass is zero, but in this case
the accumulated charge would be very small because the charge
accumulation must stop when the Coulomb force becomes equal to the
gravitational one. For massive photons the Coulomb field disappears
during $\Delta t \sim 1/\mg$ and much larger charge could be 
created outside of black holes. However, the process 
would be inefficient in the early
universe if $\mg$ is bounded by eqs. (\ref{mg-jupiter},\ref{mg-gal}). 
On the other hand, 
if the photon mass is generated by coupling (\ref{L-R}), the characteristic
rate of disappearance of electric field inside BH would
be the same as the Hubble
rate and this does not introduce any additional suppression in addition
to the standard ones present in baryogenesis, as e.g. a smallness of 
CP-violation and entropy suppression. 

The electrogenesis in the early universe depends upon many unknowns, in
particular, on the number density of the primordial black holes,
on the properties of $X$-particles, on the magnitude of CP-violation,
etc. More realistic and much better defined mechanism is the electrogenesis
in contemporary or slightly younger universe, 
which could be realized by heavy galactic black 
holes. Such black holes have been created by stellar evolution at relatively
low red-shifts, $z\sim 0-5$. They could also be created in the early
universe, even very massive up to $10^6 M_\odot$, by the mechanism
suggested in ref.~\cite{ad-js-bh}. In the second case the process 
of the electrification would proceed during much larger interval in
red-shift and could be much more efficient. In what follows we
will concentrate on more realistic and simpler
case of contemporary black holes.

It is noteworthy that electrogenesis by the present-time BH does not
require any violation of CP invariance in elementary processes.
The necessary breaking of charge symmetry is realized by the 
baryo-asymmetric background. Only protons and electrons are present,
while there are no antiprotons and positrons.

A superheavy BH in galactic center (QSO) is surrounded 
by electron-proton plasma, ionized by the quasar radiation. The
mobility of protons in plasma is much larger than that of 
electrons. This leads to a capture of protons by 
BH and repulsion of electrons out, creating a radial electric current.
Similar mechanism was considered for charging celestial bodies in
ref.~\cite{sch} in the case of the standard massless electrodynamics.
If photons are massive, the accumulated charge can be much larger,
because the electric field of the charged BH disappears with
the rate $\Gamma_{ch} \sim \mg$ and
the process does not stop when the electric charge reaches
the maximum allowed value for $m_{\gamma}=0$:
\be 
\epsilon = \frac{N_{charge}}{N_{total}} = \frac{m_p^{2}}{\alpha\, m_{Pl}^{2}}
=10^{-36}\, ,
\label{max-charge}
\ee
determined by the equality of the gravitational attraction of protons 
and the Coulomb repulsion due to electric charge of BH. Here 
$N_{total}= M_{BH}/m_p = 10^{57}(M_{BH}/M_\odot)$
is the total number of particles in BH, $N_{charge}$ is the number
of excessive protons, and $M_\odot$ is the solar mass.

The maximum rate of the charge accumulation is given by 
\be
\dot Q = \epsilon m_\gamma N_{total} \, .\nonumber
\ee
To be more accurate, one has to solve diffusion equation for protons
in the plasma around BH but we do not do that assuming that the
diffusion rate is much larger than $\Gamma_{ch}$.

The total electric current of electrons emitted by BH would be
$J=\dot Q$. Presumably the current would not be spherically 
symmetric, $J_r = J_r (r,\theta)$,  because the
propagation of electrons in the disk encounter more resistance 
than the propagation in the orthogonal to the disk direction.  
In this way jets of electrons perpendicular to the galactic plane
accelerated by Coulomb repulsion would be created,
which charge the universe. Such currents can create large scale
magnetic fields with the strength
\be
B \sim J/R_{gal} \sim 10^{-24}\, \textrm{Gauss} 
\label{magn-gal}
\ee
for ${ R_{gal}= 1}$ kpc, ${ M_{BH} = 10^6 M_\odot},$ 
and ${ m_\gamma = 1/{ kpc}}$. 
If $m_\gamma^2 =$ R, and in the 
galactic center
$ \rho \approx 10^{-10}{ g/cm^3}$, then
$ B \approx 10^{-20}\,\,$Gauss.

The observed value at galactic scale is $B\approx 10^{-6}$ Gauss 
and the galactic dynamo about $ 10^{14}$ is necessary. It is a
large amplification but still it may be possible.
The discrepancy would be diminished assuming a large coefficient 
$\xi$ in eq. (\ref{L-R}) but it looks rather unnatural.
For a review of galactic magnetic field problem see 
ref.~\cite{magn-rev}.

The total number of the excessive electrons in the outer space generated
during the life-time of a black hole would be
\be
Q_{tot} = \dot Q\, t_{gal} = 10^{34}
\label{electrons}
\ee
if we take $ m_\gamma^2 = 1/$year and $t_{gal} = 3$ Gyr.
The total number of protons in galaxy 
$N_{gal} \approx 10^{69}$, and correspondingly 
\be
Q_{tot}/N_{gal} \approx 10^{-35} > 10^{-36},
\label{q-over-n}
\ee 
that is the Coulomb repulsion created by  electrons may exceed the 
attraction by the galactic gravitational field. It is essential that 
when electrons are outside the dense galactic center, the Compton
wavelength of photon, if created by (\ref{L-R}), 
becomes huge, larger than a few Mpc. It means that at this scale
the electric or magnetic field is not Yukawa suppressed.
The kinetic energy acquired by electrons during the galactic
life-time is tiny. The Coulomb potential is only an order of magnitude
larger than the galactic gravitational potential:
\be
\Phi = \frac{M_{gal}}{m_{Pl}^2 R_{gal}} \sim 10^{-6}\, .
\label{gal-pot}
\ee
Hence the electrons would be non-relativistic and remain in the 
vicinity of the galaxies even during all universe age, 10 Gyr.

It is tempting to prescribe the accelerated universe expansion
to the Coulomb repulsion induced by the uncompensated electric 
charges of galaxies. 
However, there are several yet unanswered questions. How efficiently
electrons can transfer their momenta to a galaxy? Does it go through
direct scattering on hydrogen, on ionized matter or on galactic 
magnetic field? If the accelerated expansion is indeed enforced by
Coulomb repulsion, what happens with the angular spectrum of cosmic
microwave radiation? The position of the first peak at the point corresponding to the exactly flat
universe would be accidental and not induced by inflation. 
The total cosmological mass density in this case would be about
0.3 of the critical one. Most probably the effect of electro-repulsion 
is small but still it may be interesting to include it into data 
analysis.

To summarize, we have shown that if the mass of photon is non-vanishing
the universe must be electrically charged. The magnitude of cosmological
electric charge asymmetry is determined by the unknown value of $\mg$.
The most promising possibility to have noticeable cosmological 
implications is $\mg^2 \sim R$. In this case large scale magnetic 
fields may be explained by electric currents from
central black holes with relatively mild dynamo. Electric repulsion of 
galaxies before homogenization of charge may 
mimic the observed cosmic acceleration, though most probably the 
effect is subdominant and needs further investigation. 

\begin{center}
{\bf REFERENCES}		
\end{center}
%\begin{thebibliography}{99}

\end{document}